# A Framework for the Private Governance of Frontier Artificial Intelligence

*Dean W. Ball*[1]
Fellow, Fathom
Research Fellow, Mercatus Center
Nonresident Senior Fellow, Foundation for American Innovation

**Abstract**
This paper presents a proposal for the governance of frontier AI systems through a hybrid public-private system. Private bodies, authorized and overseen by government, provide certifications to developers of frontier AI systems on an opt-in basis. In exchange for opting in, frontier AI firms receive protections from tort liability for customer misuse of their models. Before detailing the proposal, the paper explores more commonly discussed approaches to AI governance, analyzing their strengths and flaws. It also examines the nature of frontier AI governance itself. The paper includes consideration of the political economic, institutional, legal, safety, and other merits and tradeoffs inherent in the governance system it proposes.

---

**Acknowledgements**

I am fortunate to have three institutional homes: the Mercatus Center, Fathom, and the Foundation for American Innovation. Each have nourished my work in countless ways. I wish to particularly thank Blake Pierson, Andrew Freedman, and Bri Treece from Fathom; Tyler Cowen, Ben Klutsey, and Eileen Norcross from Mercatus; and Zach Graves, Max Bodach, and Samuel Hammond from FAI.

I have also been lucky to have met a staggering number of brilliant and wise people during the course of my research. Many of these people have also been kind enough to help me develop my ideas, build my audience, and find new opportunities for policy entrepreneurship. I have been profoundly enriched by them. Sometimes, even just a single conversation was enough to strike fire.

I wish to thank: Michael Adamo, William Alexander, Mark Beall, Miles Brundage, Nathan Calvin, Martin Casado, Paul Carrier, Brian Chau, Jack Clark, Declan Garvey, Aeden Gasser-Brennan, Brendan Gorrell, Tim Hwang (and all the Good People), Mike Knoop, Daniel Kokotajlo, Seb Krier, Nathan Labenz, Nathan Lambert, Yuval Levin, Tim Lee, Keegan McBride, Max Meyer, Perry Metzger, Matthew Mittelsteadt, Richard Ngo, Chris Painter, Virginia Postrel, Nabeel Qureshi, Ketan Ramakrishnan, Santi Ruiz, Alan Rozenshtein, Cameron Schiller, Jordan Schneider, Ari Schulman, Yo Shavit, Mike Solana, Adam Thierer, and Thomas Woodside.

Thank you all, and many more.

Thank you, most of all, to my wife, Abigail, who never stopped believing in me.

**Introduction**

Frontier artificial intelligence systems will soon likely meet or exceed the performance of human beings on a wide range of cognitive tasks. They may enable the substantial or complete automation of many domains of knowledge work—everything from online marketing to advanced mathematics to AI research and development itself. The capabilities and ambitions of governments, firms, and individuals are likely to expand radically because of this new technology. So, too, will the capacity of malicious actors to cause harm.

The range of conceivable governance challenges with a technology as general purpose as *intelligence itself* is astonishingly broad. There are foreseeable risks which a wide variety of actors—from industry to government to academia to civil society—have made progress in addressing. But alongside those efforts, there is a meta-problem that needs to be solved: the development a *governance structure*. Not a solution to any discrete problem, but an articulation of the set of mechanisms, processes, and checks and balances used to solve problems themselves. This is what the economist Friedrich Hayek would have called a *nomocratic* (process-oriented) rather than a *teleocratic* (goal-oriented) structure.[2]

This dimension of frontier AI governance is somewhat more akin to the design of the Constitution, a document that codifies processes for the solving of problems and resolution of disputes without specifying any goal in particular. It is to this task that this paper is devoted.

This paper will present a framework for the private governance of frontier artificial intelligence systems. That framework will be based upon a statutory mechanism whereby developers of AI systems can opt in to oversight by private bodies whose function is to validate that the developers have met a heightened standard of care. In exchange, the developers receive protections from tort liability for *user misuse* of their systems.

The private bodies would be authorized and overseen by government. Government could authorize as many private bodies as it deems fit, allowing different private governance approaches to compete with one another on a variety of dimensions (cost, compliance verification strategies, technical standards, and the like). The proposal includes safeguards against a "race to the bottom" by regulators and industry. By operating as a voluntary regulatory regime, this system of private governance would compete, in essence, with the status quo.

For the reader who is only interested in the specifics of the proposal, Section 4 will serve you best. The remainder of the paper builds context and motivations for the proposal.

The paper will proceed as follows. The first section will begin by briefly outlining some of the foreseeable governance challenges posed by frontier AI. Then, it will consider a use-based governance system, where regulation focuses on policing illicit *uses* of AI rather than the development of AI systems themselves. While it will not argue *against* use-based regulation, it will argue both that use-based regulation can easily become onerous, and that even a well-designed use-based regulatory regime is insufficient to meet the challenges at hand. This section will also argue that governance *during the transition to advanced AI* faces meaningfully distinct constraints from governance *after* that transition has taken place.

---

[2] Hayek, Friedrich A. von. *Law, Legislation, and Liberty, Volume 1: Rules and Order.* The University of Chicago Press. 1973.

The second section will articulate why traditional government regulation or standards-setting mechanisms are unlikely to adapt well and with sufficient haste to these challenges. In doing so, this section will argue that a centralized government regulator is suboptimal for reasons of both practicality and political economy.

The third section will examine other proposed frontier AI governance frameworks: compute governance, tort liability, and international governance. It will not dismiss any of these frameworks out of hand, but instead outline their shortcomings as a governance framework *during the transition* to frontier AI and the acute stresses this transition is likely to bring with it.

The fourth section will turn to private governance of frontier AI development as a solution. It will explain the proposal sketched above in concrete detail, including mechanisms intended to lessen the likelihood of one of the failure modes coming to fruition. It will explore parameters that could be adjusted while maintaining the spirit of the proposal. And it will explain why authorizing *multiple* private governance bodies is an essential part of the proposal, while taking no position on which, if any, of these approaches to governance is best. Indeed, the objective of this paper is to argue that different approaches should be put into trial and competition with one another.

The private governance regime put forth here is intended to address a wide variety of conceivable known and unknown challenges AI will pose. But its primary intent is to address the gravest harms: loss of personal or digital property (whether by fraud, cyberattacks, etc.) and physical harm. This is why the primary incentive for developers to participate in the system is safe harbor from *tort* liability, not other forms of statutory liability such as civil rights and consumer protection laws. However, other harms, such as algorithmic bias, could be incorporated into this regime if policymakers and public believe it wise to do so. This would need to be accompanied by a concomitant safe harbor against analogous forms of liability (i.e., a private governance regime that includes algorithmic bias would need to also include protections against liability under civil rights law).

Finally, this proposal is primarily intended to cover the *development* of *software* systems. It is not a governance regime for other physical manifestations of AI, such as self-driving cars, consumer or industrial robotics, autonomous drones, and the like. The author's tentative conclusion is that existing legal frameworks are reasonably suited to governing these technologies. It is the development of infinitely scalable and replicable digital intelligences which this proposal intends to address.

# What is AI Governance?

*1.1 The Challenges at the Frontier*

For purposes of this paper, governing is the intellectual task of setting rules for how systems and resources should be used. Governing also entails additional work, such as establishing mechanisms for enforcing those rules, creating processes for resolving disputes about how the rules should apply, and ensuring accountability if someone disobeys the rules.

Frontier AI governance specifically focuses on ensuring the safety, security, and reliability of advanced AI systems. If these systems possess the profoundly novel capabilities that many industry figures expect, they will require qualitatively different governance approaches than the major technology waves of the past half-century. It is worth explicating why this is the case:

1. **AI systems are intended by their developers to match or exceed human intellectual capabilities.** This includes the ability to perform any cognitive task that a computer can perform using a computer—creating cyberattacks, conducting fraud, threatening or intimidating others, and a vast range of other activities. The potential for misuse is therefore high.
2. **AI systems, being software, are infinitely scalable and replicable**. This means that harmful activity can be scaled rapidly and used to overwhelm societal defenses.
3. **AI systems improve rapidly.** Current rates of algorithmic efficiency improvements are on the order of 400% per year, and this occurs as the cost of compute falls over time. Thus, any capability available at a high cost now can be expected to be an order of magnitude cheaper (and available to a wider range of actors) within approximately 12 months.
4. **Researchers do not fully (or even, arguably mostly) understand how AI systems work or how to control them**. This includes a mechanistic understanding of how they function, predicting their outputs, ensuring their objectives are aligned with actual human user objectives, and maintaining model robustness and security against adversarial attacks. Given that even current models are more like biological systems in their complexity than they are like deterministic software systems, and that this complexity is only likely to grow, it is likely that these problems will *never* be solved "perfectly." They are canonical examples of problems that will be solved piecemeal, through *bricolage*,[3] because they are classic examples of "muddle through" or "wicked" problems.[4]

Those seeking a *governance* structure for frontier AI development should neither seek nor expect that such a system will result in absolute control over *all* AI development. Given the sheer

---

[3] Lévi-Strauss, Claude. *The Savage Mind*. The University of Chicago Press. 1962.

For application of this concept to AI, see also Stahl et al., "Challenges of AI Alignment—Negotiating the Plurality of Values in AI Ecosystems through Bricolage," *Scientific Reports,* 2025.

[4] Madhavan, Guru. *Wicked Problems: How to Engineer a Better World*. 2024.

weight and momentum of the observations laid out above, no such outcome is likely to be achievable.

One can also argue that it is not *desirable* for any set of actors to maintain absolute control over the development of something as powerful as AI.[5] But this is almost beside the point: the current trajectory in the United States and much of the world is for this technology to develop and diffuse with very few tailored governance mechanisms. One can question whether totalitarian control of AI is wise, but that is not an option on the table. Governance, in this case, should not aim for such a level of control, but instead aim for something more modest: a modicum of order.[6]

To say that something is *governed* is to make a specific set of claims. It is not to say, in general, that it is *controlled absolutely* by the entities that govern it. Instead, governing is a specific kind of relationship between the thing that is governed and those entities that do the governing. While governance sometimes "bans" conduct, practices, or products, very often it does not. Instead, it frequently operates by creating *incentives* for compliance through both the assignment of accountability (fines, liability damages, etc.) and of benefits (legal protections, easier market access, and the like).

Given the speed with which AI development capabilities diffuse to smaller actors whose practices will be difficult for the state to make legible, and the vast range of uses to which "AI" can be applied, it is unlikely that a general-purpose governance regime is currently possible. Our hope and expectation for the near-term should be something more achievable—namely, the governance of *frontier* AI.

*1.2 The Objectives and Character of Frontier AI Governance*

Frontier AI governance has several characteristics to recommend it as the primary focus of short-term AI policy.[7] First, it concentrates policymakers on the most acute emerging risks and opportunities, allowing the state to develop a "preview" of what capabilities will be in the hands of a much wider range of actors (including nefarious actors) soon. Second, because frontier AI systems are built by a small number of firms, frontier AI development is relatively legible to the law without a massive near-term boost in state capacity. Third, given that frontier AI firms tend to lead their industry in capabilities, market share, and resources, their adoption of best practices, technical standards, and other governance mechanisms is likely to create an industry benchmark for others to follow.

---

[5] As actors with a diverse range of perspectives have, e.g.:
Senator Ted Cruz, remarks to the Special Competitive Studies Project, March 2025
Verdegem, Pieter. "Dismantling AI Capitalism: The Commons as an Alternative to the Power Concentration of Big Tech." *AI & Society,* April 2022.
Andreesen, Marc. "The Techno-Optimist Manifesto." October 2023.
Ball, Dean W. "Decentralized Training and the Fall of Compute Thresholds." *Hyperdimensional*. October 2024.

[6] A list of questions that might be considered relevant for maintaining the "modicum of order" can be found in Reuel et al. "Open Problems in Technical AI Governance," July 2024

[7] Other reasoning for a focus on the frontier is available in e.g. Anderljung et al., "Frontier AI Regulation: Managing Emerging Risks to Public Safety." November 2023.

Perhaps most importantly, though, it is at the frontier that novel and world-changing capabilities are likeliest to emerge first. Societies, including free societies based on the principles of classical liberalism, have a legitimate interest in constraining the possibilities for catastrophic misuse of technologies whose capabilities have only *just* reached the world.[8] It is at the frontier that society is likely to be shocked at what it finds (for the good and for the bad). It is at the frontier that society is likely to be least prepared. So it is at the frontier that society is wisest to construct its safeguards.

What might those safeguards include? A non-exhaustive list might include:

1. Verification of model training data for security vulnerabilities[9]
2. Research and implementation of model alignment and control methods
3. Adversarial robustness against prompt injections and model jailbreaks[10]
4. Usage monitoring for detection of illicit requests[11]
5. Providing information to the public about model characteristics, evaluations, risks, and developer risk management practices
6. Providing models to government for pre-deployment testing for national security-related risks
7. Providing information to government and the public on model usage for economic and labor market analysis

---

[8] Mill, John Stuart. *On Liberty.* 1859.
Hayek, Friedrich A. von. *The Constitution of Liberty*, Chapter 4 ("Freedom, Reason, and Tradition"). 1960.
Buchanan, James. *The Limits of Liberty: Between Anarchy and Leviathan.* 1975.
Nozick, Robert. *Anarchy, State, and Utopia.* 1974.
[9] See both Hubinger et al. "Sleeper Agents: Training Deceptive LLMs that Persist Through Safety Training." (January 2024) and Anthropic, "Simple Probes Can Catch Sleeper Agents" (April 2024) for examples of both proactive risk identification and mitigation.
[10] See Lapid, Raz et al. "Open Sesame! Universal Black Box Jailbreaking of Large Language Models" (September 2023) for research on the problem, and for proposed mitigations:
Sharma, Mrinank et al. "Constitutional Classifiers: Defending Against Universal Jailbreaks across Thousands of Hours of Red Teaming" (February 2025) and Zaremba, Wojciech et al. "Trading Inference-Time Compute for Adversarial Robustness." (January 2025)
[11] Usage monitoring is often, though not always, infeasible for developers of open-source and open-weight AI systems. Other items on this list are impossible for developers of such systems to guarantee, because having access to model weights allows malicious actors to remove model safeguards, frequently with minimal compute. The resolution to this problem is unclear. Open-source and open-weight models provide unique benefits in terms of competitiveness, facilitation of open science, and technology diffusion. Currently, the benefits of these systems at the frontier likely outweigh their marginal risks (see, for example, Kapoor et al., "On the Societal Impact of Open Foundation Models"). This could change for tomorrow's frontier. Making this decision is a difficult matter of both technocracy and politics is one of the many functions a frontier AI governance regime would serve, though it is worth stressing that the question of whether open models are desirable *at the frontier* is entirely separate from whether they are desirable *in general*.

Each of these tasks is one for which only a frontier model developer could take primary responsibility. These are the kinds of tasks a frontier AI governance system should aim to incentivize and accelerate. Most also require technical research and allocation of limited compute budgets, meaning that information sharing in these domains among frontier AI developers could conceivably create efficiencies. Information sharing of this kind by companies themselves risks running afoul of antitrust law, meaning that a third party intermediary (either government or a private governance body of some kind) is required to facilitate it; this is another ancillary benefit of a governance regime.

*After* the transition to advanced AI has gained hold, and some measure of stability has been found, an altogether different governance regime may be called for than the one designed to govern the frontier. This new regime may be more expansive or considerably lighter touch. Unfortunately, it is impossible to predict what our needs will be not only because of the uncertainty associated with how advanced AI will transform the world, but also because of uncertainty associated with how it will transform *government*, and thus the *kinds of governance* that are possible in the first place.

Many frontier AI policy proposals have focused on regulating *frontier models*. This approach typically involves creating a threshold based on characteristics of the model above which regulation of some kind is triggered.[12] The challenge with this approach is that the frontier is in constant and rapid motion; today's frontier is tomorrow's old news. Thus, a frontier threshold today will cover a far broader range of models in the not-too-distant future, eventually coming to encompass smaller firms that might lack the ability to comply with the regulations.

Even with dynamic thresholds designed to adjust over time, there are challenges to using "the AI model" as the basic conceptual unit of frontier AI governance (or indeed, any AI governance). Even today, models change rapidly, with incremental updates to their weights delivered on a near-weekly basis. Some of these are publicly announced "feature" updates, and some are more akin to "bug fixes," to use traditional software analogies.

In the recent past, we have seen the birth of the "reasoning" paradigm, which employs large-scale, reinforcement-learning-based *posttraining* on a *pretrained* language model.[13] Many

---

[12] See e.g.
Executive Office of the President, Executive Order 14110. "Safe, Secure, and Trustworthy Development and Use of Artificial Intelligence." 88 FR 75191, 2023-24283. (rescinded)

European Union Artificial Intelligence Act, Chapter V, Article 51. "Classification of General-Purpose AI Models as General-Purpose AI Models with Systemic Risk."

Safe and Secure Innovation for Frontier Artificial Intelligence Models Act (SB 1047). California Senate, 2023-2024 legislative session.

[13] The most detailed technical explanation of this approach is available in DeepSeek-AI, "DeepSeek-R1: Incentivizing Reasoning Capability in LLMs via Reinforcement Learning."
See also OpenAI, "OpenAI o1 System Card."

model-based thresholds, designed before the reasoning paradigm had entered the picture, were caught flat-footed and would not have captured these reasoning models despite the models being plainly at the frontier.[14] What's more, looking ahead, the idea of "the model" as a singular artifact could become even more outdated, with various methods of continual learning (and thus constant adjustment of weights) under investigation by researchers.[15]

All these trends push against using the model as the basis for frontier AI regulation.[16] A better approach would be to rely upon a concept that is unlikely to go out of fashion so rapidly: the corporate entity.[17] Regulating frontier AI *firms* could include not just their largest and most expensive models, but *all* models developed by the firm, whether for internal or public deployment, as well as the firm's business and research practices, internal processes, and holistic approach to safety and security.

Regulating frontier AI firms would not just be a recognition that the practices of these firms pose unique risks to society (a contestable observation, but an argument this paper supports). It would also recognize that the AI services they provide form a kind of domestic (and likely global) digital infrastructure, requiring *continual* and *dynamic* safety and security procedures. These *entities* are likely to be the primary stewards of digital intelligences which in turn undergird a nontrivial fraction of cognitive activity in the world. Their effects on culture, the economy, government, and much else will be profound, and entity-based regulation merely recognizes the unique character of these firms in particular. The governance of these firms is likely to require a *macroprudential* posture.

How to recognize *entities* as the subjects of regulation while mitigating against the possibility of regulatory capture is a crucial topic to which we will return in section four.

*1.3 The Limitations of Use-Based AI Regulation*

Before proceeding to a discussion of various approaches to frontier AI governance, it is necessary to address two questions that are often raised in AI policy discussions: should we not simply regulate malicious AI *uses*? And are not most malicious uses of AI *already illegal*?

The answer depends largely upon what is meant by "regulating AI uses." Certainly, *post hoc* enforcement of existing statutes against plainly malicious and currently illegal uses of AI (e.g.

---

fraud) are desirable. Few would dispute that this is the basis of a sound governance regime for *any* technology.

The trouble, however, is twofold. First, many use-based governance regimes are, in practice, *preemptive* rather than *post hoc* use-based regulations. Instead of enforcing the law after a harm has occurred, these regimes focus on avoiding the possibility of harm by adding on procedural compliance burdens to the (often) already extensive set of legal obligations companies face under existing statutory and common law.[18]

While there are undoubtedly *some* areas where the use of AI does merit additional, preemptive risk assessment by deployers, overreliance on such an approach risks spreading an undue compliance burden to firms throughout the economy. In general, preexisting tort and statutory liability provides an incentive for firms (particularly in heavily regulated domains of commercial activity) to exercise care prior to deployment. Regulators would be better served by developing quantitative, objective standards to which the industries they oversee should comply. Simply adding more paperwork to the equation increases an already-large compliance burden while fundamentally punting on the question of "what does a responsible deployment of AI in a given context look like?"[19]

Second, many cases of malfeasance will be more effectively and efficiently addressed by action on the developer*,* rather than the deployer, side of the value chain. For example, if agentic AI systems of the near future can be used commit fraud at scale (say, 10,000 successful cases of fraud in a day), the use-based regime would have law enforcement respond to each of those cases without, necessarily, any ramifications for the developer of the model that enabled the harm. Developers, in this world, would be under no or very little obligation to preemptively mitigate such harms, and law enforcement would be left scrambling amid a potential tsunami of novel misconduct.[20]

Without a doubt, law enforcement will have to learn to scale its own responses in a world replete with advanced AI. But placing this burden entire on the *post hoc* actions of government, and hence on the wallet of the taxpayer, is fundamentally unjust. While this paper will later argue against the use of tort liability in particular as a means of compelling frontier AI developers to internalize negative externalities of their services, there is no doubt that any sound frontier AI governance regime must compel such internalization to at least some meaningful extent.

---

[18] See, e.g.
"Concerning Consumer Protections in Interactions with Artificial Intelligence Systems," Colorado SB 205 (signed May 17, 2024)
European Union Artificial Intelligence Act, Chapter III, Article 27, "Fundamental Rights Impact Assessment for High-Risk AI Systems."
[19] For more discussion, see the "A Longer-Term Vision for AISI" section in Ball, Dean W. "On the US AI Safety Institute." *Hyperdimensional*. March 2025.
[20] Brundage et al. "The Malicious Use of Artificial Intelligence." February 2018.

Use-based regulation, including preemptive regulation, is not without its place in the overall landscape of AI governance. And *post hoc* enforcement of the law should and likely will remain the primary means of policing illicit conduct enabled by AI. But frontier developers have an important and enduring role to play in ensuring that major harms from their systems, including to uninvolved third parties (that is, neither the user nor the developer), are mitigated.

**Why Centralized Government AI Regulation is Undesirable**

*2.1 The Political Economy of Centralized AI Regulation*

Existing AI systems have already shaken the status quo in numerous ways both subtle and obvious.[21] If advanced AI systems are capable of automating significant fractions of human knowledge labor, it seems reasonable to suppose that they will challenge the status quo far more directly and viscerally.[22] In addition to this, frontier AI firms themselves could wield large amounts of power and influence over the political and regulatory system. It is therefore worth thinking carefully about the incentives of a centralized government AI regulator.

A useful lens here is that of public choice economics, which studies regulatory bodies not as abstract entities, but instead as organizations composed of individuals—bureaucrats writing and enforcing rules, lobbyists striving to influence that process, and politicians overseeing the apparatus.[23] Public choice examines the incentives of those individuals, aiming to create a sketch of what regulation is *likely* to achieve rather than what it is *intended* or *desired* to achieve.[24]

The most obvious takeaway from public choice economics is that regulated firms often shape rules in subtle ways that reflect their business interests. This is known as "rent-seeking" or "regulatory capture."[25] This process need not be a corrupt one. Often, it arises almost sociologically, through simple patterns of interaction with regulated entities and their regulators.[26] It is especially likely to arise in complex technical fields where both the measurement and mitigation of risks are high-dimensional tasks. Very often in these situations, there is a substantial information asymmetry between the regulator and the regulated firms, and it is hard for regulators *not* to defer to the expertise of the firms they regulate.[27]

---

[21] Brynjolfson, Erik and Tom Mitchell. "What Can Machine Learning Do? Workforce Implications." *Science* vol. 358, issue 6370. December 2017.
[22] Eloundou, Tyna et al. "GPTs are GPTs: An Early Look at the Labor Market Impact Potential of Large Language Models." August 2023.
[23] Buchanan, James M. and Gordon Tullock. *The Calculus of Consent: Logical Foundations of Constitutional Democracy*. 1962.
[24] Stigler, George J. "The Theory of Economic Regulation." *Bell Journal of Economics and Management Science*, Vol. 2 No. 1, Spring 1971. RAND Corporation.
[25] Krueger, Anne. O. "The Political Economy of the Rent-Seeking Society." *American Economic Review,* June 1974.
[26] Carpenter, Daniel and David Moss. *Preventing Regulatory Capture: Special Interest Influence and How to Limit It.* Cambridge University Press. 2013.

But regulatory capture by regulated industry is far from the only challenge with a centralized AI regulator. For one thing, the technology industry has been remarkably unlikely to lobby for regulatory moats against competition in recent decades.[28] But for another, there are a far greater number of special interests that could attempt to sway a centralized regulator.[29]

AI seems poised to challenge the economic interests of a wide variety of entrenched groups, including, but far from limited to: primary and higher education, the healthcare industry, financial services, lawyers, media companies, and a vast range of other white collar professional services. Together, these groups compose some of the most powerful interest groups in American politics, as well as a significant fraction of US Gross Domestic Product.[30]

These groups have a legitimate right to advocate for their interests. But empowering a centralized, general-purpose regulator of frontier AI systems risks creating a legal chokepoint, allowing any and all of these groups to lobby for regulations that hinder the velocity of AI development, the utility of AI products (for example, by requiring AI outputs that compete with the work of licensed professionals to be reviewed by a licensed human in that field[31]), and the overall diffusion of AI technology in society.[32]

The interests of any particular set of economic actors may run counter to the overall interests of society. Weighing these tradeoffs is a fundamentally *political*, rather than *technocratic* or even particularly *legal*, process. These disputes, therefore, are the primary function of legislatures rather than bureaucracies. A centralized government regulator, in addition to being ill-suited to the task, short circuits this fundamentally political process—a process which is undoubtedly to consume much of the attention of policymakers over the coming decades. America's politicians

---

[27] Laffont, Jean-Jacques and Jean Tirole. "The Politics of Government Decision-Making: A Theory of Regulatory Capture." *The Quarterly Journal of Economics*, Vol. 106, No. 4. Oxford University Press, November 1991.
[28] Indeed, large technology firms have often used existing regulatory structures to their distinct advantage rather than seeking to erect new regulatory moats. See, for example, Khan, Lina M. "Amazon's Antitrust Paradox." *The Yale Law Journal*, Volume 126, No. 3. January 2017.

"It is as if Bezos charted the company's growth by first drawing a map of antitrust laws, and then devising routes to smoothly bypass them. With its missionary zeal for consumers, Amazon has marched toward monopoly by singing the tune of contemporary antitrust." (Page 716) can and should look at these conflicts between consumer surplus and special interests directly in the face, rather than punting them to the domain of an ostensibly technical regulatory body. But a centralized government regulator makes this punting all too likely.

[29] Wilson, James Q. (Ed). *The Politics of Regulation*. 1980.
[30] Bureau of Economic Analysis. Gross Output by Industry. Accessed March 30, 2025.
[31] Section 13, "An Act Relating to Artificial Intelligence Systems," SB 199, Nevada Senate.
[32] There is ample evidence that special interest groups have lobbied to reduce the utility and development of existing, potentially competitive technologies, e.g.:
Barton, Benjamin H. and Deborah L. Rhode. "Access to Justice and Routine Legal Services: New Technologies Meet Bar Regulators." *Hastings Law Journal*, Volume 70, Issue 4.
Rauch, Daniel E. and David Schleicher. "Like Uber, but for Local Government Regulation: The Future of Local Regulation of the 'Sharing Economy.'" *The Ohio State Law Journal*, Volume 76, Issue 4. 2015.

To be clear, these political economy challenges are likely risks to *any* regulation of almost *any* technology. And they are likely to present themselves with AI no matter what regulatory approach we pursue. But they are especially acute in the centralized regulation of a general-purpose technology. Thus, additional care must be taken to mitigate against adverse political economy outcomes.

In the discussion to follow on the private governance strategy proposed by this paper, we will explore how competitive private regulatory bodies (imperfectly) mitigate against many of these challenges. But it is worth briefly observing that many of these political economy problems stem from, or are at least worsened by, the fact that traditional regulation envisions government as possessing a *monopoly* on the task of regulation itself. If one accepts the premise that regulation in the abstract, and frontier AI governance in particular, are *pro-social* goods, then it should also follow that the production function—the *structure*—which produces those goods deserves scrutiny. The central question is this: how many goods of similar societal value do we entrust to a *monopoly*? The virtue of competition is thoroughly rooted in the American psyche and in bipartisan political rhetoric. But in the case of regulation and technocratic governance, that virtue is thrown by virtually all scholars, politicians, and other observers out of the window. It is assumed that regulation *must* be the domain of a monopoly provider—the government. This paper, and the proposal it offers, seeks to challenge that basic assumption.[33]

These theoretical and structural concerns are cause enough to question the efficacy of a centralized AI regulator. Unfortunately, there are numerous practical challenges as well.

*2.2 The Practical Barriers to Government AI Regulation*

As a general matter, any regulation (public or private) of AI—even just of frontier AI—presents numerous practical challenges. The technology evolves rapidly. It can be difficult to define precisely what it is.[34] Frontier AI regulation is technically daunting and requires answering questions that currently have no good answers (and in some cases, may never; see the prior discussion on the wicked problems of frontier AI governance). It is difficult to hire the human experts necessary to execute on frontier AI governance.[35]

---

[33] I am here indebted first and foremost to the work of Gillian Hadfield and Jack Clark in their paper "Regulatory Markets: The Future of AI Governance." (April 2023)

[34] Even at the frontier, definitions are difficult. See the preceding discussion in Section 1.2

[35] Note that this challenge *may* be worse in the public sector than in the private sector. We have good reason, for example, to believe that fixed government pay scales and lengthy hiring timelines can be an impediment to hiring relevant expertise within bureaucracies. Yet it is also the case that, for example, the United Kingdom and United States AI Safety Institutes have, by most accounts, staffed themselves ably with experts willing to

All of these are challenges that any AI governance body will face. There are reasons to believe that competing private governance bodies will face them more ably than a monopoly government regulator will. These reasons will be discussed in depth in a later section of this paper. For the time being, though, we should consider these challenges to be the table stakes in this particular game of poker. The task of this section is to consider what *additional* drawbacks government regulators bring to the table. resources, trust)

First is the fact that AI governance is likely to require the aggressive and creative adoption of emerging technologies, not the least of which is frontier AI systems themselves. It is hard to imagine how one could meaningfully govern a new information and cognitive technology without the use of that same technology. How could one effectively govern, for example, the internet, without also using the internet themselves?

Yet government agencies tend to lag in technology adoption for a complex myriad of cultural, bureaucratic, and legal reasons.[36] This is a widely recognized problem in American public administration, and progress is conceivable. Progress might even lead to governments modernizing data infrastructure, fully adopting cloud computing, and successfully diffusing other technologies of the last decade. But even if that progress comes to fruition, it is unlikely that government will become a leading-edge adopter of emerging technologies, or an *innovator* in their use. At the very least, it is a risky bet to take with something as crucial as frontier AI governance. On the other hand, it is generally accepted that private firms tend to outpace governing in the adoption of emerging technologies.[37]

Second, there is the fact that the political process often fails to allocate capital in ways suited to meet the needs of administrative agencies. The United States federal government,[38] as well as many state and local governments,[39] face significant debt burdens that make funding ambitious new government initiatives difficult.

The most relevant government effort proximate to frontier AI governance is the export controls placed on AI computing hardware and semiconductor manufacturing equipment by the Biden Administration beginning in October 2022.[40] Despite frequently winning bipartisan support in Congress, the enforcing agency—the Bureau of Industry and Security within the United States Department of Commerce—has not received the funding it needs to conduct this function

---

face those obstacles in the interest of public service. Thus I do not include human capital acquisition in the list of challenges *uniquely* facing government.
[36] Pahlka, Jennifer. *Recoding America: Why Government is Failing in the Digital Age and How We Can Do Better*. 2023.
[37] Hinkley, Sara. "Technology in the Public Sector and the Future of Government Work." University of California-Berkeley Labor Center. January 2023.
[38] United States Senate Joint Economic Committee. "National Debt Growth." March 2025.
[39] United States Census Bureau. "Annual Survey of State Government Finances." October 2024.
[40] Bureau of Industry and Security, Department of Commerce. "Commerce Implements New Export Controls on Advanced Computing and Semiconductor Manufacturing Items to the People's Republic of China (PRC)." October 7, 2022.

properly.[41] It uses legacy information technology systems that render it far less capable of performing its job even with limited operational support[42]—itself an indicator of underfunding. Not only is the additional funding to fulfill this bipartisan national security function not appear to be forthcoming—there is a chance that BIS' funding could be *reduced* in the coming fiscal year.[43]

We therefore have good reason to believe that even if government sets up a well-structured and properly scoped frontier AI governance function, and even if that body is given the authority it needs to acquire technology and human talent as quickly and flexibly as necessary, and even if support for this function is bipartisan, that Congress *still* might systematically under-fund the entity in question. Meanwhile, we know that the philanthropic sector alone has ample interest in supporting AI governance initiatives, and have strong reason to believe that a private frontier AI governance body (or several) could be amply funded with private dollars.

Finally, there is the issue of *trust*. As discussed above, frontier AI governance is likely to require the sharing of data, best practices, and other information between frontier AI firms. These firms are unlikely to engage in such information sharing purely among one another due to potential risk of antitrust investigations. Thus, some sort of intermediary is required to facilitate this exchange. Given that the information in question is very likely to be commercially sensitive, this intermediary will need to be well-trusted by all parties. Firms would need to trust that the information would not be used against them by an aggressive regulator (either the AI regulator themselves or an adjacent government regulator that could obtain access). A demonstrably neutral private governance body is, at the margin, better positioned to serve this role—and indeed, a nascent frontier AI governance body called the Frontier Model Forum is already doing this.[44]

Frontier AI governance is challenging enough on its own. Centralized government regulation faces a wide range of both political economy and practicality drawbacks that make it less likely to succeed in the near term when compared to alternative governance arrangements. But private governance is only one of those alternatives. Let us now examine alternative *public* AI governance structures.

---

[41] Hammond, Samuel. "Supporting the Bureau of Industry and Security." Testimony delivered to U.S. House Committee on Appropriations, Subcommittee on Commerce, Justice, Science, and Related Agencies. May 2024.
[42] Office of Congressman Jason Crow. "Reps. Crow and Kean Introduce Bill to Modernize the Bureau of Industry and Security's IT Systems." August 2024.
[43] Selinger, Marc. "BIS Could See 10.5% Funding Cut in 2025, Senator Says." *Export Compliance Daily*. March 2025.
[44] Frontier Model Forum (FMF). "FMF Announces First-of-its-Kind Information-Sharing Agreement." March 2025.

**Exploring Alternatives to Centralized Frontier AI Governance**

*3.1 Compute Governance*

Compute governance seeks to manage frontier AI through the hardware on which frontier AI systems are trained and deployed—most commonly graphics processing units (GPUs), tensor processing units (TPUs), or, for model inference, application-specific integrated circuits (ASICs). The export controls on advanced AI hardware, referenced above, are an example of compute governance in practice. Compute governance can also include measures to ensure visibility and trackability, such as "on-chip governance"—dedicated coprocessors embedded into larger chips that track the device's location, security status, and other characteristics.[45] It could even involve more ambitious steps, such as the creation of a unilateral or multi-lateral, government-backed "Compute Reserve" that would operate akin to a central bank, adjusting levels of compute available to the global marketplace based on the desirability of doing so to policymakers. In short, compute governance spans a wide range of potential policies, with the only commonality being their focus on hardware rather than software.[46]

There is no frontier AI without frontier compute. And compute is a physical good, making it much more legible to the policymaking apparatus than AI models, which are infinitely replicable software. Indeed, the highest-end AI computing hardware, often made on leading-edge semiconductor nodes, is among the most difficult-to-produce goods mankind has ever devised. All these factors lend themselves positively to using compute as the primary means of governing frontier AI.

Compute-based governance has been used thus far as a means of keeping AI capabilities out of the hands of America's adversaries, most notably China. Despite noteworthy flaws, the Biden Administration's export controls have worked reasonably well, seeming to have blunted China's ability to maintain a frontier AI ecosystem. In the long run, the controls could backfire significantly. The imposition of the controls appears to have caused China to redouble its commitment to aggressively invest in catching up to the United States on semiconductor design and manufacturing.[47] If China eventually does develop its own leading-edge semiconductor ecosystem, US export controls on American products will no longer function as intended.

The decision to implement the export controls, however wise it might have been, also created path dependency in American policymaking. The inevitable result of government controls on the movement of desirable physical goods is that those goods will be transported illicitly or quasi-illicitly. There is substantial evidence for both illicit[48] and quasi-illicit[49] movement of high-end computing hardware after the export controls took effect. The result of this is that, within approximately two years, the Biden Administration's export controls ratcheted up from a "small yard, high fence" approach of restricting Chinese access to a narrow range of products to, as the Biden Diffusion Framework put it, "regulat[ing] the global diffusion of the most advanced artificial intelligence models and large clusters of advanced computing integrated circuits."[50] Implementing the "small yard, high fence" strategy, it turned out, required regulating the construction of large data centers, export of advanced compute, and use of frontier models in every country on Earth.

---

[45] Aarne, Onni et al. "Secure, Governable Chips: Using On-Chip Mechanisms to Manage National Security Risks from AI & Advanced Computing." Center for a New American Security. January 2024.

[46] A broader survey of the range of options in compute governance—and a well-articulated argument in favor of pursuing compute governance—is available in Sastry et al. "Computing Power and the Governance of Artificial Intelligence." Centre for the Governance of AI. February 2024.

[47] Triolo, Paul. "A New Era for the Chinese Semiconductor Industry: Beijing Responds to Export Controls." *American Affairs*. Spring 2024.

And despite these downsides, even some of the leading proponents of export controls tend to agree that the controls will not deny China (and with enough time, any adversary) the ability to develop frontier models, as DeepSeek's v3 and r1 models partially demonstrated. Instead, the proponents argue, export controls are intended to deny our adversaries AI *ecosystems*—the economy-wide benefits of AI that can only come from widespread use of compute.[51] This logic is sensible, but it does not change the fact that, if frontier AI is indeed a kind of weapon that our adversaries wish to use against America or the West, export controls are unlikely to stop them.

Thus, even under the best of circumstances, compute export controls have significant tradeoffs. The tradeoffs may have been worth it, and the policy as a whole may have been wise. But any additional compute governance mechanisms—such as the on-chip governance discussed briefly above—are likely to only add to the incentive for adversary countries to develop an alternative computing ecosystem. Once that ecosystem is in place—and this is an inevitable outcome—compute governance as a foreign policy tool is no longer viable.

The limitations of compute governance are even more apparent in the creation of a *domestic* governance regime—the objective of this paper. Compute governance so far has been used to *limit* or to deny outright advanced AI capabilities to American adversaries. Even on-chip governance ideas have been put forth primary to advance these broader objectives of limitation or denial. But it would be senseless, of course, to try to limit or deny compute to the domestic economy.

---

[48] Grunewald, Erich. "AI Chip Smuggling into China: Potential Paths, Quantities, and Countermeasures." Institute for AI Policy and Strategy. October 2023.
[49] From the Bureau of Industry and Security: "BIS is also concerned regarding the potential for China to use IaaS solutions to undermine the effectiveness of the October 7 IFR controls and continues to evaluate how it may approach this through a regulatory response."
"Implementation of Additional Export Controls: Certain Advanced Computing Items; Supercomputer and Semiconductor End Use; Updates and Corrections," 88 FR 73458, 2023-23055, pgs. 73458-73517. October 2023.

[50] Bureau of Industry and Security. "Framework for Artificial Intelligence Diffusion." 90 FR 4544, 2025-00636, pgs. 4554-4584. January 2025.
[51] Lee, Timothy B., Dean W. Ball, and Lennart Heim. "Lennart Heim on the AI Diffusion Rule." *AI Summer*. January 2025.

In principle, compute governance could be used as a domestic tool to ensure that malicious uses of models are curtailed at the lowest possible level of the AI supply chain—the physical hardware itself. On-chip governance, for example, could be used to examine the computations running on the primary part of the AI chip to ensure that it is not a malicious usage of an AI model. This would take pressure off model providers to do such usage monitoring themselves. In particular, developers of open-source and open-weight frontier models would be freed from any obligation to monitor their models' usage, which is otherwise impossible with such models.

The significant and unshakeable tradeoff is that this on-chip governance mechanism would amount to, at the very least, a government-mandated security vulnerability in all frontier AI computing hardware,[52] and at worst, a backdoor giving government the option to surveil all AI computing performed using those chips.

In addition to being a violation of privacy and creating a potentially enormous security vulnerability in American computing infrastructure, such a mechanism could very well incentivize AI deployment into a kind of underground market. It is already possible to run sophisticated models locally on consumer and prosumer hardware. Increasingly, it is even possible to *train* modestly sophisticated models on decentralized networks of geographically distributed computing hardware.[53]

There are some domestic policy objectives that compute governance and similar approaches (such as data center-level regulations) *can* achieve. For example, some scholars have recently suggested trading relaxed federal permitting for large scale computing clusters in exchange for the resultant data center adhering to stringent cybersecurity requirements.[54] This is a pro-social and eminently achievable goal.

But for any broader domestic frontier AI governance system, with the goals described above in Section 1.2, compute governance simply is not suitable.

*3.2 Tort Liability and Insurance*

A "tort," coming originally from the Latin *tortum* ("injustice," "twisted") is a wrongful act committed by one person that results in injury—physical harm, loss of property, or both—to another. In the United States, it is primarily a creature of state, rather than federal, law (the vast majority of tort cases are filed in state courts[55]). This also means that tort law varies considerably from state to state, though private governance bodies such as the American Law Institute exist to harmonize the standards for tort litigation to some extent.[56]

---

[52] The recent, and quite possibly ongoing, Chinese infiltration of American telecommunications systems was enabled at least in part by a government-mandated backdoor (in the Communications Assistance for Law Enforcement Act) to permit law enforcement wiretaps.
Krouse, Sarah et al. "US Wiretap Systems Targeted in China-Linked Hack." *The Wall Street Journal*. October 5, 2024.
[53] Jaghouar, Sami et al. "INTELLECT-1 Technical Report." December 2024.
[54] Datta, Arnab and Tim Fist. "Compute in America: A Policy Playbook." Institute for Progress, February 2025.

Tort liability of various forms has been proposed by many as a tool to incentivize frontier AI developers to take heightened care with regard to the safety and security of their AI systems.[57] It is almost certainly the case that tort liability of some form or another *already applies* to frontier AI developers for at least some harms; the specifics will be determined either piecemeal by courts as cases present themselves, or top-down by legislatures.[58] The most prominent proposed AI laws in the United States have relied upon tort liability (negligence in particular) as a means of enforcement.[59] In other words, tort liability, broadly conceived, is America's current default governance mechanism for frontier AI development. It is worthwhile, therefore, to take a step back and examine the system as a whole.

The tort system as it exists today in the United States originates in, but is distinct from, English common law—law made by judges in the resolution of individual cases rather than by legislators in the passage of statutes. Most importantly, US tort law underwent a transformation in the mid-20th century from a purely legal mechanism to a tool of economic policymaking. This was an era when society was still contending with the rise of the managerial corporation, mass-produced goods, and (relatively) new inventions like the automobile, modern pharmaceuticals, and an immense diversity of other goods that posed novel dangers to individuals. At the time, liability for the corporations that made these goods was primarily determined through contract warranty claims (which often contained broad disclaimers of liability) or negligence liability, which at the time was exceedingly difficult for plaintiffs (victims) to establish.

Legal scholars, armed with concepts from economics, sought to cause corporations to internalize their negative externalities, and they sought to use the tort system to accomplish this fundamental public policy objective. The scholars and jurists who led this movement began to apply strict liability (liability regardless of how much care was exercised by the manufacturer) to products, and at the same time to soften the standard for proving negligence on the part of the seller of a service or good.

---

[55] Bureau of Justice Statistics, "Tort Cases in Large Counties." United States Department of Justice, 1992.
[56] See e.g. *A Concise Restatement of Torts, Third Edition*. 2013.
[57] See e.g.:
Weil, Gabriel. "Tort Law as a Tool for Mitigating Catastrophic Risk from AI." January 2024.
Choi, Bryan H. "Negligence Liability for AI Developers." *Lawfare*. September 2024.
Sharkey, Catherine. "Products Liability for Artificial Intelligence." *Lawfare*. September 2024.
[58] Ramakrishnan, Ketan et al. "US Tort Liability for Large-Scale Artificial Intelligence Damages: A Primer for Developers and Policymakers." RAND Corporation. August 2024.
[59] Safe and Secure Innovation for Frontier Artificial Intelligence Models Act (SB 1047). California Senate, 2023-2024 legislative session.

These standards were codified in 1964 by the American Law Institute—the above-mentioned private governance body—in its *Restatement (Second) of Torts*. Within a few years, they were recognized by most states across the country as the law of the land. Section 402(A) of the second *Restatement*, which most directly codified this change to tort law, has been described as "the most cited section of any [American Law Institute] Restatement," and gave birth to an immense body of subsequent case law.[60] [61] While tort law originates in the ancient (or at least old) common law tradition, the system as it exists today was very much the result of a deliberate and top-down effort.

The consequences of this effort were profound. Some argue that it was necessary to incentivize firms to internalize the negative effects their products had on society, and to invest in superior safety practices. Others argue that it creates a "lottery," in which companies are held responsible essentially at random for harms alleged by sympathetic victims, regardless of how much culpability the company really had for the harm.[62]

One observation that is near-indisputable is that tort liability as it existed by the 1980s caused a panic in the insurance market. Manufacturers of vaccines, airplanes, sports equipment, practitioners of a diversity of medical services, providers of day care, municipalities, and a great many other entities found themselves unable to afford—or sometimes even to receive—insurance, owing to the immense uncertainty created by the liability system.[63]

Substantial reforms have been made to tort systems throughout the United States since the 1980s and proceeding through the 2000s, primarily to deal with what many saw as excesses of the original system.

The vast majority of tort cases result in either dismissal or settlement; vanishingly few proceed to trial. This means that almost none of the harms contemplated by tort law result in a court-mediated fact-finding procedure, and that almost none of them result in precedent to guide other courts (a key feature and benefit of the common law). When cases do proceed to trial, courts

---

[60] Vandall, Frank J. "Constructing Products Liability: Reforms in Theory and Procedure." *Villanova Law Review,* Volume 48, Issue 3. 2003.
[61] It is worth noting that Section 402(A), and all of the *Restatement (Second) of Torts*, have been superseded by the *Restatement (Third) of Torts: Products Liability* in 1997.
[62] For reading in favor of the mid-century tort law changes, it is best to look to its architects:
*Escola v. Coca Cola Bottling Co.*, 24 Cal.2d 453, 150 P.2d 436 (Cal. 1944) (Traynor, J., concurring)
*Greenman v. Yuba Power Products, Inc.*, 59 Cal.2d 57, 27 Cal. Rptr. 697, 377 P.2d 897 (Cal. 1963)
Calabresi, Guido. *The Costs of Accidents*. 1970.

For reading opposed:
Huber, Peter W. *Liability: The Revolution and Its Consequences*. 1990.
Kagan, Robert A. *Adversarial Legalism: The American Way of Law*. 2001.
Epstein, Richard A. "Products Liability as an Insurance Market." *The Journal of Legal Studies*, Volume 14, Number 3. December 1985.
[63] Priest, George L. "The Current Insurance Crisis and Modern Tort Law." *The Yale Law Journal*, Volume 96, Number 7. June 1987.

(judges and juries selected at random) have been known to award strikingly disparate damages for similar harms, which raises questions about how well the system prices the negative externalities it is supposed to cause firms to internalize.

Moreover, while estimates vary, most agree that roughly 50% of the monetary compensation awarded in tort litigation is paid to victims, with the rest going to lawyers' fees, administrative costs, and other ancillary expenses.[64] Other victim compensation schemes, such as the Vaccine Injury Compensation Program,[65] which is an administrative alternative to tort liability, are estimated to deliver 85% of damages to victims. Thus, tort liability does not always deliver on its promise of providing just compensation to victims of corporate misfeasance or incentivizing firms to internalize their negative externalities.

On the other hand, the more recent reforms to the tort system do seem to have resulted in a more predictable, coherent system. Many states have capped non-economic and punitive damages, modified standards for negligence, heightened standards for scientific evidence, and provided defenses for companies whose products were approved or otherwise complied with federal regulator standards.[66] While complex and multi-dimensional, these reforms generally bring tort law in line with its original intended public-policy purpose.

Common law solutions to emerging technologies also have the distinct advantage of allowing the law to adapt iteratively to harms as they manifest themselves in the real world.[67] This is often preferable to politicians writing preemptive laws meant to address imagined and hypothetical risks.[68]

Common law and tort liability therefore have a vital role to play in the governance of frontier AI and many other emerging technologies likely to be associated with it (consumer robotics, autonomous vehicles and drones, etc.). Throughout the economy, individuals and businesses are sure to use AI systems, both digital and physical, to engage in a kaleidoscopic and unpredictable variety of conduct. Some of this conduct will cause harms. Tort liability is reasonably well suited to contend with these harms.

The more specific question considered by this paper, however, is whether tort liability should play a predominant (or any) role in the governance of frontier AI *development*. It is here that tort liability shows its weaknesses most acutely.

---

[64] Congressional Budget Office. "The Economics of US Tort Liability: A Primer," page xi. 2003.
[65] Itself a federal workaround created to make vaccines viable to manufacture and sell in the United States after tort liability made such activity much more difficult.
[66] McQuillan, Lawrence J. and Hovannes Abramyan. US Tort Liability Index: 2010 Report. Pacific Research Institute. 2010.
[67] Shavell, Steven. "Liability for Harm Versus Regulation of Safety." *The Journal of Legal Studies*, Volume 13, Number 2. June 1984.
[68] Calabresi, Guido. *A Common Law for the Age of Statutes*. Harvard University Press. 1982.

The question principally centers on misuse by customers and third-party harm resulting from that misuse. No one disputes that if an employee slips and falls in the OpenAI headquarters, the company should face potential premises liability. More seriously, few would dispute that if an OpenAI model, in testing for autonomy capabilities, were to exfiltrate itself from its secure testing environment and begin defrauding unsuspecting people on the internet (for example), OpenAI could and likely should face tort liability for this.

But such harms are not the focus of liability-related frontier AI governance proposals like SB 1047. Instead, those proposals focus on the use of negligence liability on frontier AI developers to police harms caused to third parties by customer misuse. That is, developers would be expected to adhere to a standard of care that would be defined by judges and juries in the context of tort litigation. If developers met or exceeded this standard of care, they would not face liability. If they were found to have acted beneath this standard, they would face liability.

In contemplating the potential consequences of this legal standard being applied to frontier AI development, it is important to remember that AI is a *platform* or *foundational* technology. That is, AI will eventually constitute a new substrate of technology embedded across civilization, in turn enabling all (or practically all) social, economic, scientific, and technological activity that takes place above it. In addition to being a consumer and business technology, it is likely to become a kind of new computing primitive—a new foundation on top of which all software is built.[69] Its structural role will be similar to that of other platform technologies such as the internet, electricity, telecommunications networks, and (in an earlier era) the railroad.

*All* of the providers of these platform technologies enjoy significant or near-total protections from tort liability arising from misuse. If I use the phone system to defraud thousands of people, the victims have no recourse to sue the phone company. An electricity company can be held liable (often in strict liability) if its transmission lines fail due to improper maintenance and cause a fire;[70] but it would be incoherent to speak of suing the power company because one of their customers *misused* electricity.[71] Similarly, if a Google AI data center were to explode due to negligent management of the facility, damaging a home on an adjacent property, most would intuitively agree that Google should be held liable.

America does not grant tort liability protections to providers of platform technologies lightly, or out of some sense of libertarian largesse. Indeed, many of these industries are among the most highly regulated in American economic life. Instead, these liability protections recognize a basic fact: the technologies in question are so fundamental and so immensely wide-ranging in their conceivable uses (and misuses) that their harms (often grave and life-threatening harms) are better addressed through statute and regulation than through tort litigation. Forcing providers of these technologies to internalize "their" negative externalities, if those externalities include user misuse, is akin in the limit to forcing them to internalize the negative externalities of civilization itself. And this, of course, no firm can do.

---

[69] A speculative but enlightening discussion on this topic is available from OpenAI Co-Founder Andrej Karpathy in his lecture "Intro to Large Language Models."
[70] In fact, even electricity *outages* are generally not actionable under private tort causes. See *Waldon v. Arizona Public Service Co.*, No. 14-55076 (9th Cir. 2016).
[71] Though in principle, every crime committed with a computer is a misuse of electricity.

Ironically, State Senator Scott Wiener epitomized the dangers of an approach like this just one year after he introduced SB 1047 with another bill called SB 222,[72] which would allow tort lawsuits against fossil fuel companies for the January 2025 Los Angeles wildfires, on the theory that the burning of fossil fuels caused climate change, which in turn caused the wildfires and the resultant property damage. The logical flaws in this chain of causation are numerous,[73] but they also reveal something deeper: an enormous range of legitimate harms can be tied back to the firms that provide or manufacture such fundamental technologies.

The fact that the technology under examination here is the development of *mechanized intelligence* makes liability a particularly perilous tool (if, that is, we care about that technology continuing to be developed by American companies). Liability doctrines rely upon the question of whether a harm was *reasonably foreseeable* to the manufacturer, or in this case, to the developer. A *railroad* may be fundamental to civilization, but its range of foreseeable (and even unforeseeable) misuses and harms is considerably more limited than *intelligence*—the ability to recognize patterns and solve problems. Given that intelligence undergirds literally all human action, it is hard to demarcate what, exactly would *not* be reasonably foreseeable to the company that developed human-level mechanized intelligence.

Imagine a bizarre world where, somehow, the American tort liability system exists, but humans are still basically apes—we have no higher cognitive faculties. Then one day, a person finds a spring with miraculous water that gives those who drink it full human intelligence. The person fences off the spring and begins selling the water to the other, still-pre-rational *homo sapiens*. Should the person who stumbled on the spring have *liability exposure for everything everyone does with human intelligence for all time*? Not, this paper contends, if we want there to be a cognition industry for very long.

One does not need to cast judgments on tort liability *in general*. One simply needs to concur that it is a suboptimal way of governing user misuse of platform technologies, a fact which is reflected by a century of American technology policy.

There are more practical considerations as well. *Any* liability regime for frontier AI development would likely amount to *de facto* strict liability (liability that obtains regardless of how much care the developer exercised) for open-source and open-weight models. It will *always* be possible for developers to exercise more robust safety and security policies on closed-weight models than their open-weight counterparts.

---

[72] Climate Disasters: Civil Actions (SB 222). California Senate, 2025-2026 legislative session.
[73] With many arguing that the state's own insurance policies, public land mismanagement, and electrical infrastructure maintenance practices were far more likely causes. See e.g. Handmer, Casey. "The Los Angeles Wildfires Are Self-Inflicted."
Powell, Lars et al. "Rethinking Prop 103's Approach to Insurance Regulation." International Center for Law & Economics. November 2023.

Then there is the fact that the tort system, famously, incentivizes legal action against the party with the greatest financial resources, and not always the party that is most culpable for the harm in question.[74] In cases involving AI systems, this is often likely to be the frontier model developer. Take, for example, the recent Chinese AI agent product called Manus. This was a computer-using agent that attracted attention for its novel capabilities but also concern about its lack of safeguards. The system was marketed (primarily to Westerners) by a Chinese startup called Monica AI and featured models made by Alibaba (a large Chinese technology firm) and the US frontier AI developer Anthropic.

If a firm had deployed this system in a way that caused a plausibly tortious harm, the easiest legal target by far is Anthropic, regardless of how culpable Anthropic had been in creating the harm. These dynamics would no doubt structure the incentives of potential litigants, given that the time and cost of entering into tort litigation is substantial. The system incentivizes would-be litigants to file suits most likely to earn compensatory relief, and not necessarily most likely to assign negative externalities to the most appropriate actor.

Finally, there is the simple fact that a frontier AI governance system based on tort litigation exposes the industry—and thereby the entire economy—to a great deal of uncertainty. If policymakers choose to let the status quo persist (in which tort liability of some form likely obtains), or if they explicitly opt for a liability-first AI governance regime (with laws such as SB 1047), then the first substantive AI governance standards are likely to be written by judges and juries with no particular expertise on the topic.

Not only will the authors be randomly chosen. So too will their subject matter, defined as tort litigation outcomes are by the peculiar facts of the cases at hand. This is a substantial risk. It may be a risk America could bear in normal conditions, but one must not forget that AI is being developed in the context of a heated geopolitical competition with China and, plausibly, other countries. In AI development, every month matters. One case, one court, one bad ruling could easily be enough to take the American AI industry off of its current development trajectory and cadence for far more than one month. Too, it could indelibly shape the future development of the technology with judicial doctrine that is fundamentally unknowable in advance. This is an exceptionally high level of risk to tolerate.

There can be little argument that tort liability will play a meaningful and positive role in the overall governance of AI (especially at the deployment, or use, layer) and of the many technologies AI will enable. But for frontier AI development, tort liability is ill-suited for reasons both structural and practical.

---

[74] Schwartz, Victor E. et al. "[Deep Pocket Jurisprudence: Where Tort Law Should Draw the Line](#)." *Oklahoma Law Review*, Volume 70, Number 2. 2018.

*3.3 International Governance*

Many academics and industry leaders, from Nobel Prize Winner Geoffrey Hinton[75] to OpenAI CEO Sam Altman[76] and others,[77] have advocated for some kind of global governance system to oversee advanced AI. While these esteemed figures may be correct that AI systems may one day reach a level of capability so extreme as to require such an arrangement, there is currently no path to achieving it.

Global affairs today are characterized by increasing strife, competition, and, indeed, the fraying of international governance institutions of all kinds. The United States and China, the two leading countries in the development of artificial intelligence, are engaged in a protracted struggle over a wide variety of geopolitical conflicts; technical and governance standards for emerging technologies are one of them. The political tenor of our age, unfortunately, does not lend itself well to ambitious global coordination and governance.

Any consideration of international AI governance must also grapple seriously with the tension between controlling advanced AI and recognizing the sovereignty of all countries on Earth. AI will be a general-purpose, civilizational fundament. It is the right of each country to choose how they wish to use the technology, within exceptionally wide bounds limited only by truly catastrophic or existential harms. How such bounds would be policed without the implementation of a global surveillance system over all uses of AI remains to be seen. In other words, it remains to be seen whether any kind of robust and durable global governance regime is even possible, let alone politically achievable.

While this paper does not dispute the prudence of diplomatic efforts related to international AI governance, it argues that America's efforts toward frontier AI governance are best channeled toward domestic efforts. Ideally, those domestic efforts would be designed to be easily exportable to other countries should the opportunity and mutual desire arise. There may be narrow windows of opportunity in the future for international AI governance cooperation, and those windows are most likely to be seized by the United States if the country already has well-developed, demonstrably functional governance practices in place. Because private governance organizations operate more readily across international lines than do governments, this paper argues that the governance approach most likely to facilitate international collaboration in the long term is private governance.

---

[75] Hinton, Geoffrey. 2024 Nobel Prize in Physics Banquet Speech.
[76] Sam Altman remarks to United States Senate Judiciary Committee, Subcommittee on Privacy, Technology, and the Law. *Tech Policy Press*.
[77] Bengio, Yoshua et al. "Managing Extreme AI Risks Amid Rapid Progress." *Science*. May 2024.

**Private Governance: An Adaptable Framework**

*4.1 An Outline of the Proposal*

We have already considered traditional regulation, compute governance, governance by tort liability, and international governance, and found all four solutions to be wanting in various important ways.

This paper proposes a hybrid public-private governance framework for frontier AI development. The basic structure is as follows:

1. A legislature authorizes a government commission to license *private* AI standards-setting and regulatory organizations. These licenses are granted to organizations with technical and legal credibility, and with demonstrated independence from industry.

2. AI developers, in turn, can **opt in** to receiving certifications from those private bodies. The certifications verify that an AI developer meets technical standards for security and safety published by the private body. The private body periodically (once per year) conducts audits of each developer to ensure that they are, in fact, meeting the standards.

3. In exchange for being certified, AI developers receive safe harbor from all tort liability related to misuse by others that results in tortious harm.

4. The authorizing government body periodically audits and re-licenses each private regulatory body.

5. If an AI developer behaves in a way that would legally qualify as reckless, deceitful, or grossly negligent, the safe harbor does not apply.

6. The private governance body can revoke an AI developer's safe harbor protections for non-compliance.

7. The authorizing government body has the power to revoke a private regulator's license if they are found to have behaved negligently (for example, ignoring instances of developer non-compliance).

The proposal's primary incentive for joining is the safe harbor from tort liability for model misuse by others. Because of this, the authorizing government body and the private governance organizations would ideally be focused on mitigating tort-related risks (physical injury, loss of or damage to physical or digital property) and achieving an elevated standard of care.

The authorizing government body can (and ideally, should) authorize *multiple* private governance organizations. This has several benefits. First, it permits competition among private governance bodies, mitigating against the tendency of regulation to gradually increase in complexity and compliance cost over time.

Second, competition among private governance organizations allows innovation and experimentation in the design and implementation of governance. One core thesis undergirding this proposal is that advanced AI will itself be a breakthrough governance technology, enabling new ways for regulatory organizations to design standards (say, by the automated creation of AI model evaluations) and enforce them (for example, through automated monitoring facilitated by tightly bound AI agents).

Third, the authorization of multiple private governance organizations avoids the one-size-fits-all nature of technocratic regulation when it is promulgated by a singular government entity. Instead, it allows institutional entrepreneurs to carve out "niches" in the "governance marketplace." Perhaps there will be a need for startups marketing agents, based on heavily modified foundation models, to have their own private governance organization. Open-source and open-weight models could be subject to standards of their own, reflecting the distinct differences in practical governance solutions with closed-source models. New markets that develop over time, such as robotics or biological foundation models, could be served with dedicated governance organizations, rather than having to be squeezed into some pre-existing regulatory design.

Because the majority of legislative efforts on AI have taken place at the state government level so far,[78] this proposal is designed to accommodate implementation by state or federal governments. Specifically, the private governance bodies are intended to operate across state lines. Thus, state governments can enact frontier AI governance (satisfying a key demand of voters) while avoiding the creation of a state-by-state compliance patchwork.

Ideally, however, this proposal would be enacted by the federal government, which can ensure that one nationwide legal standard obtains (as opposed to some states with tort liability safe harbor and private governance, and some states with the status quo or alternative governance regimes), and has the most relevant expertise to certify the private governance bodies responsibly.

Additional detail on the core structural elements of the proposal follows.

*4.2 The Authorizing Government Body*

The government body that authorizes private governance organizations is the key public oversight mechanism in this proposal, and the source of all authority vested in the private governance organizations. It is the heart of what makes this proposal a hybrid, rather than a purely private, governance arrangement.

The membership, powers, and constraints on the authorizing government body are therefore key. As envisioned by this proposal, the authorizing government body would be a multi-member commission, with fixed members representing government agencies with special expertise in AI (for example, the Director of NIST or of the US AI Safety Institute, the Director of the White House Office of Science and Technology Policy, etc.) as well as members appointed by Congress and the President (or, in the event of a state implementation, the Governor and the state legislature).

---

[78] Multistate.AI. "AI Legislation." March 2025.

A state government implementation would also need different fixed committee members, since there is no analog to the science and technology-focused positions mentioned above in states. These fixed roles could include the Attorney General, leaders of in-state public research universities, and others. In both state and federal implementations, the appointed members would ideally come from outside of government, and represent, for example, the academic and research communities.

The central power granted to this body is the ability to review and approve applications from private governance bodies, investigatory powers to determine if a private governance body has behaved negligently, and the capacity to revoke a governance body's certification license. In the ideal implementation of this proposal, the application review process would be narrowly focused on the technical capability of the applicant to conduct the certifications it envisions. The priority of all parties—the private governance organizations and the authorizing government body alike—would be on the mitigation of plausible tort-related harm, most notably loss of property and physical harm.

All other considerations should be explicitly disallowed by any statutory implementation of this proposal, unless the proposal is modified to provide liability protections in analogous legal domains. For example, if legislators wish for this proposal to be adapted to mitigating algorithmic bias, protections from liability under civil rights law would need to be added or to replace the tort liability protections. As designed, this proposal is focused on the mitigation of tort harm-related risks, and its incentive for joining is intended to be analogous to these intended risks.

To mitigate against mission creep and political interference, the authorizing government body's powers should remain both significant and narrow. It should retain an absolute right to certify and decertify private governance organizations, and to conduct relevant investigations into the conduct of the organizations and the firms they certify. But the authorizing government body should not, for example, be granted broad rulemaking authority.

The authorizing government body has the power to revoke a private governance organization's license in the event that negligence or misfeasance is discovered. Doing so would eliminate tort liability protections for *all* companies covered by that private governance organization. This mechanism is intended to prevent a "race to the bottom," where frontier AI developers flock to whichever regulatory body is perceived as most lenient.

Say, for example, that Company A and Company B are both frontier AI model developers seeking certification from Private Regulator C, who is known to be the most lenient in the governance market. If Company B's models, poorly overseen by Private Regulator C, cause a major harm, the authorizing government body can revoke Private Regulator C's license, which will *also* remove liability protections from Company A. Thus, Companies A and B alike have an incentive to avoid piling into a knowingly lax private governance body. This a blunt instrument, but it is designed to ensure that the authorizing government body only revokes licenses in the event of serious misconduct.

*4.3 The Private Governance Bodies*

The private governance bodies perform the vast bulk of the work envisioned by this proposal. Their function is to develop technical standards, model evaluations, auditing and monitoring mechanisms, and other governance mechanisms to ensure that frontier AI firms and models meet an elevated standard of care.

The proposal envisions minimal regulations or restrictions on *how* these private bodies would go about their tasks, beyond the oversight (both *ad hoc* and scheduled, as mandated by the periodic license reviews) by the authorizing government body. This lack of prescriptive rules is very much by design. As mentioned above, a motivating thesis for this proposal is that advanced AI is likely to transform the task of governance *itself*. Generalist AI agents are likely to radically increase the cognitive labor capabilities of small teams of humans, as well as, in the long term, the optimal configuration of organizations themselves.[79]

The creation of new, private organizations to experiment with different regulatory approaches and institutional designs, and the competition between those private organizations, is intended to facilitate the societal discovery of new ways to govern with advanced AI.

As Gillian Hadfield and Jack Clark[80] put it in their "Regulatory Markets" paper, which is among the largest sources of inspiration for this proposal:

"Private regulators could use conventional regulatory tools—establishing written requirements, monitoring for compliance with those requirements, and penalizing violations—but we also expect that private regulators would develop new regulatory technologies... Such technologies could themselves be data-intensive and deploy AI methods."

And:

"In any given domain, multiple regulators are licensed so that they compete to provide regulatory services to targets. Targets must choose a regulator but they have the capacity to choose, and switch, regulators. They do so by comparing across regulators in terms of the cost and efficiency of the services provided by regulators."[81]

---

[79] This would not be new in the history of general-purpose technologies. For example, the railroad and telegraph created novel demands on firms that gave rise to the modern managerial corporation. A similar transformation, into some unpredictable new form, is likely to occur with advanced AI. For more on the development of the modern managerial corporation, see:
Chandler Jr., Alfred D. *The Visible Hand: The Managerial Revolution in American Business*. 1977.
[80] Clark is a co-founder of Anthropic, a frontier AI development firm that would be a covered entity in the proposal described here.

The objective of the regulatory market envisioned in this proposal is to protect against tort-related harms *before* they happen. In that sense, the regulatory structure proposed is preemptive. Over time, as more real-world data on usage, risks, and harms become available, lawmakers could set specific targets by which to measure the performance of private regulators (for example, the number of incidents resulting in damage to property or physical injury stemming from models developed by covered firms).

Because this system of regulatory markets exists alongside the existing legal status quo, there are two levels of competition envisioned. One is the competition within the regulatory market, as already described. The other is the competition *of the regulatory market as a whole* with the status quo. Some believe that the existing system of tort liability is not just a more effective, but also a lighter-touch, way of achieving harm reduction from frontier AI systems. This contention should itself be put to a kind of market test, by allowing firms to assess the relative merits of each system for themselves. This proposal enables such a test to be conducted.

One key challenge with the private governance organizations proposed here will be ensuring their independence from industry while also allowing the organizations to recruit the requisite human capital. It is probable that at least some of the expertise needed to run a well-functioning private governance body would come from the frontier AI industry. These individuals may well have direct or indirect financial exposure to one or more frontier AI development firms.

Potential solutions to this problem include divestment requirements, a mandate that employees with recent frontier AI industry experience may not serve on private governance body boards, and a statutorily defined maximum ratio of private governance body staff with industry experience. The routine audits envisioned by the proposal could also be tailored to monitor this issue in particular.

4.4 Liability Protections

The problems with tort liability for user misuse of platform technologies like frontier AI are discussed above in Section 3.2. Rather than reiterating those problems here, this section will explore the mechanics of the envisioned liability protections as well as some of the parameters policymakers could consider adjusting while preserving the basic spirit of the proposal.

First, the liability protections apply only to liability in tort, and only to harms arising from user misuse of a frontier model. They do not apply to standard corporate tort liability exposure (premises liability, workplace harms, or other forms of traditional corporate liability), nor do they apply to any *statutory* liability companies may face (civil rights, consumer protection, environmental regulations, and other large bodies of law). Tort liability exposure is also preserved for harms stemming from *first-party* use of frontier models (e.g. an internal deployment of a model that exfiltrates itself from its secure environment and commits tortious acts against third parties).

---

[81] Hadfield, Gillian and Jack Clark. "Regulatory Markets: The Future of AI Governance." 2023.

Second, while a full safe harbor from user-misuse-related tort liability is ideal for creating an incentive for firms to participate, it is not the only option. The safe harbor could be modified such that it only applies to harms below a certain objective count, for example incidents involving property damage of less than $500 million[82] or deaths below a threshold. Above that threshold, the protection could shift to a rebuttable presumption of reasonable care on the part of the developer, or protections could be eliminated altogether.

Third, the rebuttable presumption could also be expanded to cover *all* tort liability claims. That is, rather than a full safe harbor, the liability protections could take the form of a rebuttable presumption against all claims. This would soften the protections, quite possibly to the point that some number of industry actors would decline to participate in the regulatory market. But in all likelihood, the broader spirit of the proposal would be maintained, especially if it were implemented with a rebuttable presumption at the federal level.

*4.5 Structural and Technical Considerations*

From the perspective of institutional economics, this proposal has four key advantages over traditional forms of regulation:

1. **Lower transaction costs:** This proposal lowers the transaction costs associated with frontier AI governance on the theory that private transacting bodies (covered firms and private governance bodies) can define, monitor, and enforce rules among themselves more cheaply than traditional government regulatory bodies.[83]
2. **Information asymmetries:** While information asymmetries will likely persist between frontier AI firms and any governance body, these asymmetries will likely be lower between a frontier AI firm and a private governance body with a tailor-made structure, specialized expertise, and more sophisticated ability to use technology to facilitate AI governance.
3. **Heterogeneity:** Because the AI industry is diverse, a flexible private system will be better able to serve a variety of governance needs, as opposed to a one-size-fits-all traditional regulatory solution.
4. **Reputation mechanisms:** Frontier AI firms are likely to rely heavily upon their reputations to successfully market large-scale knowledge work automation to corporate customers, especially those in regulated industries. By providing a quasi-government certification to AI developers who participate successfully in the regulatory market, this proposal creates a reputational mechanism that may be far more valuable to AI industry actors than a standard government fine.[84]

---

[82] Though estimating damages is difficult in the absence of litigation, which this proposal is intended to avoid. If a damages threshold were to be implemented, some neutral third party would need to be specified in statute to evaluate damages prior to litigation.
[83] Williamson, Oliver E. *The Economic Institutions of Capitalism*. 1985.

Though the proposal described here is imperfect (like *all* governance regimes), it meaningfully advances upon the status quo of traditional regulation in all ways outlined above.

**Conclusion**

This paper has attempted to grapple seriously with the concept of frontier AI governance. It has tried to elucidate what frontier AI governance is likely to mean in practice, and what even a well-implemented governance system can realistically be expected to achieve. No perfect order is imagined here, but neither is perfect control by the state or any other authority. Instead, this proposal offers modest order in the form of protection from the most dire conceivable harms from frontier AI. The conception of frontier AI governance offered here takes all three of political economy, liberty, and major AI risks seriously.

It then navigates one-by-one through the major forms of proposed frontier AI governance: traditional government regulation, compute governance, governance by tort liability, and international governance. It finds each of these wanting or lacking in important ways while dismissing none of them altogether. But ultimately, this paper concludes that something new will be needed to achieve sound frontier AI governance in short order.

The private governance system outlined includes safeguards against both corruption and over-regulation. It is designed to mitigate only the gravest potential harms from frontier AI, while leaving other harms to the prudent enforcement of existing law, or to the passage of new statutes. Beyond ensuring the mitigation of harms, the priority of this proposal is to incentivize institutional innovation in American governance. The author anticipates that lessons learned and experience gained from the implementation of this proposal will have numerous applications to other domains of American statecraft.

Ultimately, no governance system can be wisely put into practice without virtuous leadership and prudent enforcement. That is true for this proposal as well. Though it is designed with safeguards, it could well fall prey to the many perils facing any governance system: corruption, undue risk aversion, torpor, and the proliferation of needless complexity. Avoiding these pitfalls is often a problem of *culture*, not a problem susceptible to clever technocratic tweaks. The culture of American statecraft has seen better days. By trying something new, in the spirit of experimentation and discovery, with the gravity of the task before us in mind, perhaps culture, too, can be renewed.

---

[84] Tirole, Jean. "[A Theory of Collective Reputations (with Applications to the Persistence of Corruption and to Firm Quality](about:blank)." *The Review of Economic Studies,* Volume 63, Number 1. January 1996.